\documentclass[showpacs,preprintnumbers,amsmath,amssymb]{revtex4}

\usepackage{graphicx}
\usepackage{dcolumn}
\usepackage{bm}

\usepackage{amsmath}
\usepackage{latexsym,amssymb,graphicx}
\usepackage[mathscr]{eucal}

\begin{document}


\title{Gravitational-wave Detection With Matter-wave Interferometers\\
Based On Standing Light Waves}

\author{Dongfeng Gao\textsuperscript{1,2}, Peng Ju\textsuperscript{1,2,3}, Baocheng Zhang\textsuperscript{1,2}, Mingsheng Zhan\textsuperscript{1,2,} }

\altaffiliation{Email: mszhan@wipm.ac.cn}
\vskip 0.5cm
\affiliation{1 State Key Laboratory of Magnetic Resonance and Atomic and Molecular Physics, Wuhan Institute of Physics and Mathematics, Chinese Academy of Sciences - Wuhan National Laboratory for Optoelectronics, Wuhan 430071, China\\
2 Center for Cold Atom Physics, Chinese Academy of Sciences, Wuhan 430071, China\\
3 Graduate University of the Chinese Academy of Sciences, Beijing 100049, China
}

\date{\today}

\begin{abstract}
We study the possibility of detecting gravitational-waves with matter-wave interferometers, where atom beams are split, deflected and recombined totally by standing light waves. Our calculation shows that the phase shift is dominated by terms proportional to the time derivative of the gravitational wave amplitude. Taking into account future improvements on current technologies, it is promising to build a matter-wave interferometer detector with desired sensitivity.
\end{abstract}

\pacs{04.80.Nn, 04.80.-y, 95.55.Ym}
\maketitle

\section{Introduction}

The existence of gravitational waves is one of the most important predictions of Einstein's general theory of relativity. The indirect evidence comes from the long term observation of the binary pulsar PSR 1913+16 \cite{taylor1975,taylor1994}. Since 1970s, many efforts have been made to find direct evidences. However, it becomes more and more clear that detecting gravitational waves is really a great challenge to the experimentalists. Two types of detection technologies have been developed so far \cite{ju2000,aufmuth2004}. Resonant mass detectors are the first type of detectors. They were constructed in several laboratories. Some are still running: ALLEGRO in USA, AURIGA and  NAUTILUS in Italy, and EXPLORER at CERN. In the meantime, another type of detectors based on laser interferometers are studied and built: LIGO in USA, VIRGO in Italy, GEO600 in Germany, and TAMA300 in Japan. These present-day detectors can only reach the marginal sensitivity to detect gravitational waves from our own galaxy. To reach higher sensitivity and wider operating frequency, both types of detectors inevitably have to grow in size, and become more expensive to build. Therefore, it is of great worth to study alternative detector designs that can have comparable sensitivity, but with smaller size and less expensive.

Actually, the rapid technological advances in atom interferometry provide us a new possibility. Due to their high sensitivity, atom interferometers have already been used in many precision measurements, such as the measurement of Newton's constant G \cite{bertoldi2006}, the test of the equivalence principle \cite{dimopoulos2007}, and the breakdown of the $1/r^2$ law at small length scales \cite{ferrari2006}. One can find more details on atom interferometers in many review papers, such as \cite{cronin2009}. The theoretical investigation of detecting gravitational waves with matter-wave interferometry was started by many authors \cite{linet1976,stodolsky1979,anandan1982,borde1983}. In recent years, possible experimental schemes based on atom interferometers were extensively studied \cite{chiao2004,roura2006,foffa2006,delva2006,tino2007,kasevich2008,ambrosio2009}. These schemes use either the superposition of two different momentum states of the atom \cite{tino2007,kasevich2008}, or the particle-wave duality property of the atom \cite{chiao2004,roura2006,foffa2006}.

In Chiao and Speliotopoulos' scheme \cite{chiao2004}, atom beams are split and recombined by 32nm material gratings, and are deflected by crystal mirrors. They called their scheme, MIGO, the matter-wave interferometric gravitational-wave observatory. The limitation of their scheme is that only Helium atoms can be used due to the current ability on making crystal mirrors. It is preferable to use heavier atoms for detecting gravitational waves. On the other hand, it was experimentally known that diffraction gratings of light have more advantages than material gratings. Atom interferometers based on standing light waves were successfully constructed a long time ago \cite{rasel1995,giltner1995}.

In this paper, we study the possibility of detecting gravitational waves using the matter-wave interferometer based on Bragg scattering from standing light waves. Using Bragg scattering, we can split, reflect and recombine a variety of atom beams. It is also easier to align standing light waves than material gratings. In Bragg scattering, the atoms always remain in a single state so that the atomic phase is not sensitive to external fields and fluctuations in laser beams. The rest of the paper is organized as follows. In section II, we calculate the phase shift due to an incoming gravitational wave. The difference to laser interferometer detectors can be clearly seen from our result. In section III, we give a description of our proposed scheme, and analyze main sources of noise that in principle limit the sensitivity of our detector. Comments and conclusions will be given in section IV.

Throughout this paper, we will keep the Planck unit $\hbar$ explicitly, while adopting the usual notation for the speed of light $c=1$. Indices are lowered and raised by the Minkowski metric $\eta_{\mu\nu}$.

\section{Calculation Of The Phase Shift}

\begin{figure}[h]
\includegraphics{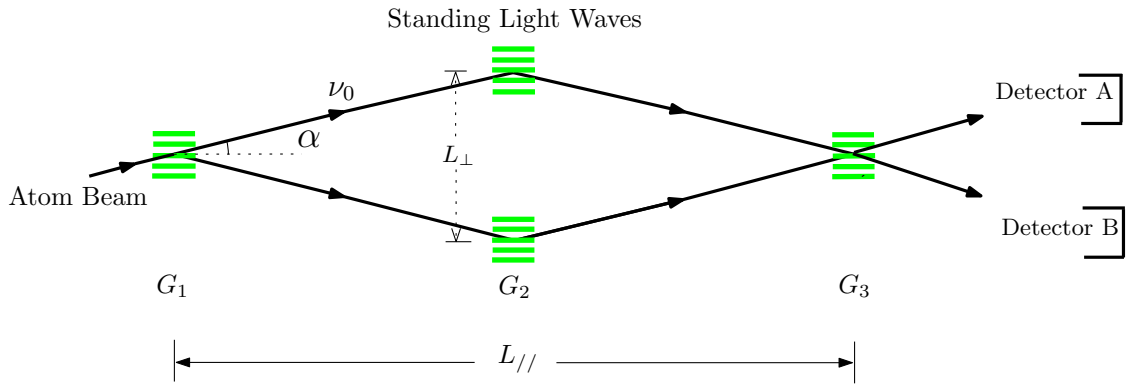}
\caption{Schematic diagram of our proposed configuration}
\label{fig1}
\end{figure}

To make the calculation clear, a schematic diagram of our configuration is depicted in FIG. \ref{fig1}. The length and width of the interferometer are denoted by $L_{//}$ and $L_{\perp}$, respectively. Atoms from the beam source with velocity $v_0$ are split into two beams by angle $2\alpha$. The two beams are then reflected, recombined and finally flow into atom detectors.

Suppose incoming gravitational waves propagate along the direction orthogonal to the plane of the interferometer, which is taken to be the $z=0$ plane. In the transverse-traceless gauge \cite{misner1973}, the metric for this linearized gravitational wave can be written as
\begin{equation}
ds^2=-dt^2 + [\delta_{ij}+h_{ij}(t-z)]dX^idX^j,
\end{equation}
where $z\equiv X^3$, $h^i_i=0$, $\partial^ih_{ij}=0$ and $i, j=1,2,3$. In literature, the metric with $h_{11}=-h_{22}=h\sin(2\pi f t +\phi_0)$ is called the $h_+$-polarization, and the one with $h_{12}=h_{21}=h\sin(2\pi f t +\phi_0)$ is called the $h_\times$-polarization.

 As our measurement is taken in laboratory, it is convenient to work in another coordinate system, related to the (t, $X^i$)-coordinates by a transformation
\begin{eqnarray*}
t &=& t,\\
x^i&=& X^i+\frac{1}{2}h^i_jX^j.
\end{eqnarray*}
The metric is then rewritten into
\begin{equation}
ds^2=-dt^2+\delta_{ij}dx^idx^j+\dot{h}_{ij}(t-z)[dz-dt]x^idx^j+ \mathcal{O}(h_{ij}^2),
\end{equation}
where terms of quadratic or higher order in $h_{ij}$ have been
neglected. In the above transformation, the condition that $\dot{h}_{ij}(t) x^{i}$ and
$h_{ij}$ are of the same order is assumed. This is necessary for the new coordinates to be consistent with the linearized treatment of the gravitational-wave metric. In other words, the condition amounts to require that the full size of the gravitational-wave detector should be much smaller than the wavelength of the gravitational wave.

It is easy to write down the geodesic equations for particles confined in the $z=0$ plane in the (t, $x^i$)-coordinates,
\begin{equation}
\frac{d^2x^i}{dt^2}=\frac{1}{2}\ddot{h}^i_jx^j+\mathcal{O}(h^2_{ij},v^2h_{ij}).
\end{equation}
One can check that the physical distances in the $z=0$ plane are not affected by gravitational waves. Therefore, this coordinate system is called the rigid frame in \cite{roura2006}. In this frame, the Lagrangian for a non-relativistic atom becomes
\begin{equation}
L(x^i,\dot{x}^i)=\frac{m}{2}(\dot{x}^i\dot{x}_i-\dot{h}_{ij}\dot{x}^ix^j-2).
\label{lag}
\end{equation}
For a given path, the phase at the final point can be calculated as
\begin{equation*}
\Phi_f=\Phi_i+\frac{1}{\hbar}S_{i\rightarrow j},
\end{equation*}
where $S_{i\rightarrow j}$ is the action from the initial to the final point, calculated from the Lagrangian (\ref{lag}).

Before the incoming of gravitational waves, atoms travel along the classical trajectories,
\begin{equation}
x_0(t)=v_0^x t \,, \,\,\,\,\,\,\,\,\,\,    y_0(t)= \left\{
\begin{array}{ll}
v_0^y t \,\,\,\,\,& {\rm for} \,\,\,\, 0<t<T/2 \\
v_0^y (T-t) \,\,\,\,\, & {\rm for}\,\,\,\, T/2<t<T \, ,
\end{array}\right.
\end{equation}
where $v_0^x=v_0 \cos \alpha$, $T=L_{//}/v_0^x$, and $v_0^y=\pm v_0 \sin \alpha$, depending on the path. In the present of gravitational waves, both the atom's trajectory and velocity will be perturbed, which can be written as $\vec{x}=\vec{x}_0+\delta\vec{x} $ and $\vec{v}=\vec{v}_0+\vec{w}$. To solve the geodesic equations for atoms, we need to know the junction conditions at the middle standing light waves. The natural conditions are that the geodesic should be continuous and the velocity should satisfy the rigid condition. That is, we have the following conditions
\begin{equation}
\delta\vec{x}|_{t\rightarrow T^-/2}=\delta\vec{x}|_{t\rightarrow T^+/2} \,\,\,\,\,\, {\rm and} \,\,\,\,\,\, \vec{w}|_{t\rightarrow T^-/2}+\vec{w}|_{t\rightarrow T^+/2}=0.
\end{equation}

Using the same derivation as in \cite{foffa2006}, the phase for each arm is found to be
\begin{equation}
\frac{\hbar}{m}\Delta_{up,\, down}\Phi=\int^T_0\{-\frac{(v_0^y)^2}{v_0^x}w^x+v_0^yw^y-\frac{1}{2}\dot{h}_{ij}v_0^ix_0^j\}|_{up,\, down},
\end{equation}
where $\Delta_{up, \, down}\Phi$ means the phase change in the up and down arms, respectively.
Because the $h_+$-polarization affects the up and down arms in the same way, our scheme is only sensitive to the $h_\times$-polarization. The phase shift is obtained by taking the difference of phase changes in the two arms,
\begin{eqnarray}
\nonumber\frac{\hbar}{m}\Delta\Phi &=& \pi f h v_0^2 T^2 \sin\alpha \tan\alpha \cos(\pi f T+\phi_0)\\
    \nonumber        & & - h v_0^2 T \sin \alpha \tan \alpha (1-\sin \alpha)\sin(\pi f T) \cos(\pi f T+\phi_0)\\
            & & + h v_0^2 T \sin \alpha \tan \alpha (1-\sin \alpha)(1-\cos(\pi f T)) \sin(\pi f T+\phi_0).
            \label{phase}
\end{eqnarray}
Note that $\Delta\Phi$ oscillates in time, because $\phi_0$ (the phase of the gravitational wave at the time of the first grating) is oscillatory. Let us discuss the result in more detail. Define
\begin{eqnarray*}
\Delta\Phi_1&=& 2\pi m A f h \tan\alpha \cos(\pi f T+\phi_0)/(\hbar\cos \alpha),\\
\Delta\Phi_2&=& \frac{m}{\hbar} h v_0^2 T \sin \alpha \tan \alpha \sin(\pi f T+\phi_0) =2hL_{\perp}k_{dB}\tan\alpha \sin(\pi f T+\phi_0),
\end{eqnarray*}
where $A=\frac{1}{2}v_0^2T^2\sin \alpha\cos\alpha$ is the area enclosed by the two arms, and $k_{dB}=m v_0/\hbar$ is the de Broglie wavenumber. In other words, $\Delta\Phi_1$ is the contribution involving the derivative of the gravitational wave amplitude, and $\Delta\Phi_2$ is the contribution involving the gravitational wave amplitude. We consider three different cases.
\begin{itemize}
\item \noindent When $f T \ll 1$,
$$\Delta\Phi \sim \Delta\Phi_1 \sin\alpha.$$
The phase shift is proportional to $\Delta\Phi_1$, and is very small since $\alpha$ is small.

\item \noindent When $f T \sim 1$,
$$\Delta\Phi \sim \Delta\Phi_1 + \Delta\Phi_2.$$
The two different contributions can be comparable.

\item \noindent When $f T \gg 1$,
$$\Delta\Phi \sim \Delta\Phi_1.$$
\end{itemize}
The feature of our result is that the phase shift contains terms from two different contributions, which have been discussed separately before \cite{chiao2004,roura2006,foffa2006,delva2006}. In this sense, our derivation is more complete. Especially, in high frequency regime ($\gg 1/T$), the contribution involving the derivative of the gravitational wave amplitude dominates the phase shift.

This phase shift can be measured by detecting the oscillation in the intensities of atom beams
going into detectors A and B. By a similar calculation to the one in \cite{champenois1999}, the intensities of atom beams detected by idealized detectors A and B are found to be
\begin{eqnarray}
I_{A}&=&\frac{I_{0}}{2}(1+\cos((\mathbf{r}_{21}-\mathbf{r}_{32})\cdot \mathbf{k}_{g}
+\Delta\Phi)), \nonumber\\
I_{B}&=&\frac{I_{0}}{2}(1-\cos((\mathbf{r}_{21}-\mathbf{r}_{32})\cdot \mathbf{k}_{g}
+\Delta\Phi)),
\end{eqnarray}
where $\mathbf{r}_{21}=\mathbf{r}_{2}-\mathbf{r}_{1}$,
$\mathbf{r}_{32}=\mathbf{r}_{3}-\mathbf{r}_{2}$, $\mathbf{r}_{i}$'s denote the
positions of light gratings. $\mathbf{k}_{g}$ is the reciprocal vector associated with
each light grating, which is assumed to be the same for all light gratings. Its norm is $||\mathbf{k}_{g}||=2\pi/a$, where $a$ is the period of light gratings. $I_{0}$ is the total intensity of the atom beam from the source. In order to maximize the detection sensitivity,
 we set $(\mathbf{r}_{21}-\mathbf{r}_{32})\cdot \mathbf{k}_{g}=\pi / 2$. Then it follows that
\begin{eqnarray}
I_{A}&=&\frac{I_{0}}{2}(1-\sin(\Delta\Phi)) \approx \frac{I_{0}}{2}(1-\Delta\Phi), \nonumber\\
I_{B}&=&\frac{I_{0}}{2}(1+\sin(\Delta\Phi)) \approx \frac{I_{0}}{2}(1+\Delta\Phi).
\end{eqnarray}
Clearly, the contrast between the atom beam intensities at the two detectors will also help to improve the reliability of the detection.

\section{Design Considerations and Noise Analysis}
\subsection{Description of our proposed configuration}

The schematic diagram was already shown in FIG. \ref{fig1}. The size of the interferometer is proposed to be $L_{//} = 200{\rm m}$ and $L_{\perp} = 1{\rm m}$. The interferometer regions must be in high vacuum about $10^{-7}$ Pa in order to keep the coherence of atom beams. At this pressure and room temperature, the vacuum contains about $n \sim 3\times 10^{13}{\rm m}^{-3}$ molecules with an average velocity of $v\sim 500{\rm m\cdot s^{-1}}$. The average collision time between atom beams and air molecules is $\frac{1}{n \sigma v}\sim 80{\rm s}$, where the typical value of cross section $\sigma$ is about $10^{-18}{\rm s^{-2}}$. So the atom beams can keep coherence for about 1s.

The atom beam is emitted from a supersonic source, with velocity $v_0=1000{\rm m\cdot s^{-1}}$. It can be well collimated by 2D optical molasses. The distribution of the transverse velocity of atoms can be controlled to meet the requirement of gravitational-wave detection. Four standing light waves are generated by laser beams from the same laser source. By adjusting the intensities of standing light waves, the first and fourth ones can function as $50\%-50\%$ beam splitters, and the two in the middle can function as beam mirrors. Interference fringes can be formed in both output atom beams of the interferometer. Either beam can be used for detection because they are complementary to each other.

To yield a good signal-to-noise ratio in our scheme, the supersonic source should have high intensity. For example, people have produced Argon beams with high flow rate $\mathcal{R}\sim {\rm 10^{19} atoms/s}$ \cite{leroy1969}. To scatter the Ar atom by angle $2\alpha=10^{-2}{\rm rad}$, we need a momentum transfer of order $10^2 \hbar k$. Currently, momentum transfer of $24\hbar k$ has been reported \cite{chu2007}. An order-of-10 improvement on large momentum transfer technology is needed. To split, reflect and recombine Ar beams of so high flow rate, the estimated laser power in our scheme is of order $10^2$watts. This high laser power can be achieved by using the power amplifier, as in LIGO \cite{ligo2009}.

Taking the above values into eq. (\ref{phase}), one can find that $\Delta \Phi =2\times10^{9}f h$. In astrophysics, there exist many interesting sources of gravitational waves, such as compact binary coalescences (CBCs) of neutron stars or black holes \cite{thorne2002}. The gravitational-wave emitting from CBCs will last for over tens of seconds with frequency in the $10-10^3$Hz range, and amplitude $h\leq 10^{-21}$. Our detector would produce a phase shift $\Delta\Phi \sim 10^{-11}-10^{-9}{\rm rad}$. Whether this phase signal can be detected depends on various sources of noise in our scheme, which can have more or less effects on the detection ability of the detector. A thorough analysis of noise is necessary for the detector design, which will be reported in our future work. Here, we focus on shot noise and seismic noise.

\subsection{Shot Noise}

\begin{figure}[h]
\includegraphics[scale=0.7]{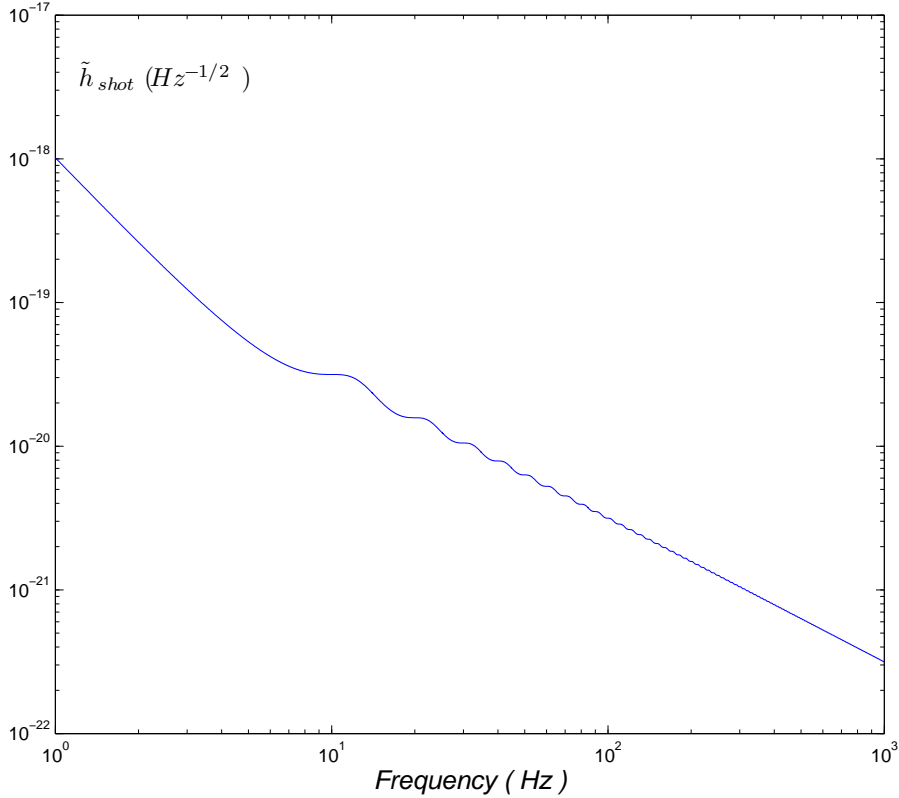}
\caption{Amplitude spectral density curve for shot noise}
\label{fig2}
\end{figure}

Shot noise is the fundamental limitation on the sensitivity of any atom interferometer. The fluctuation in the phase shift measurement is inversely proportional to the square root of the total number of atoms,
\begin{equation}
\Delta\Phi_{shot}\sim \frac{1}{\sqrt{N_0}}=\frac{1}{\sqrt{\mathcal{R}t_a}},
\end{equation}
where $N_0$ is the total number of atoms, and $t_a$ is the experimental time. This determines the minimal detectable gravitational-wave signal of the detector. To make contact with other literature on gravitational-wave detection, it is useful to write down the corresponding amplitude spectral density, which is independent of the experimental time. The formula is found to be
\begin{eqnarray}
   \tilde{h}_{shot} (f)=\frac{2\hbar }{m|C(f)|\sqrt{\mathcal{R}}}
\end{eqnarray}
where
\begin{eqnarray}
C(f)&=&[-\pi v_0^2 T^2 \sin \alpha \tan \alpha\sin(\pi f T)f +v_0^2 T \sin\alpha \tan\alpha(1-\sin \alpha) \sin^2(\pi f T)  \nonumber\\
&~~~&+v_0^2 T \sin\alpha\tan\alpha(1-\sin\alpha)(1-\cos(\pi f T))\cos(\pi f T)]  \nonumber\\
&&+i[\pi v_0^2 T^2 \sin \alpha \tan \alpha\cos(\pi f T)f
-v_0^2 T \sin\alpha \tan\alpha(1-\sin \alpha) \sin(\pi f T)\cos(\pi f T)  \nonumber\\
&~~~&+v_0^2 T \sin\alpha\tan\alpha(1-\sin\alpha)(1-\cos(\pi f T))\sin(\pi f T)].
\end{eqnarray}
Using the proposed scheme parameters, the amplitude spectral density curve is drawn in FIG.\ref{fig2}. The curve clearly shows that the sensitivity of our scheme gets better at higher frequencies.

\subsection{Seismic Noise}
Seismic noise is another important factor that will seriously affect the detection ability of the gravitational-wave detector.
Typical seismic noise levels are \cite{saulson1994}
\begin{eqnarray}
\tilde{x}_{seismic}(f)=\left\{\begin{array}{cc}10^{-9}{\rm m/\sqrt{Hz}}~ &  \,\,\,\,\,\,\,\,\,\,\,\,\,\,\,\,\,  {\rm for} \,\,\,\,\,\, 1Hz\leq f \leq 10Hz
\\ (10Hz/f)^2\cdot 10^{-9}{\rm m/\sqrt{Hz}}~ &  {\rm for} \,\,\,\,\,\,\,\, f > 10Hz
\end{array}\right.
\end{eqnarray}

In our scheme, the seismic noise has two main effects on the detection. One effect is that it will change the phase shift $\Delta\Phi$. However, this effect is very
tiny, for example, at the frequency of 100Hz, the seismic noise is about $10^{-11}{\rm m/\sqrt{Hz}}$. The interferometer area will change about $\Delta A=10^{-9}{\rm m^2/\sqrt{Hz}}$ by this seismic noise. Then the phase shift will change about $10^{-20}{\rm rad}$ for $t_a= 10{\rm s}$, which is really small.

The other effect is that the seismic noise will break the condition,
 $(\mathbf{r}_{21}-\mathbf{r}_{32})\cdot \mathbf{k}_{g}=\pi / 2$, we assumed before. If we don't adopt any vibration isolation system, the corresponding amplitude spectral density is
\begin{eqnarray}
\tilde{h}_{seismic} (f)&=&|\frac{2\pi}{a}\cdot \frac{\hbar}{mC(f)}| \sqrt{4S_{seismic}(f)}\\ \nonumber
               &=&|\frac{2\pi}{a}\cdot \frac{\hbar}{mC(f)}|\cdot 2 ~\tilde{x}_{seismic}(f).
\end{eqnarray}

\begin{figure}[h]
\includegraphics[scale=0.7]{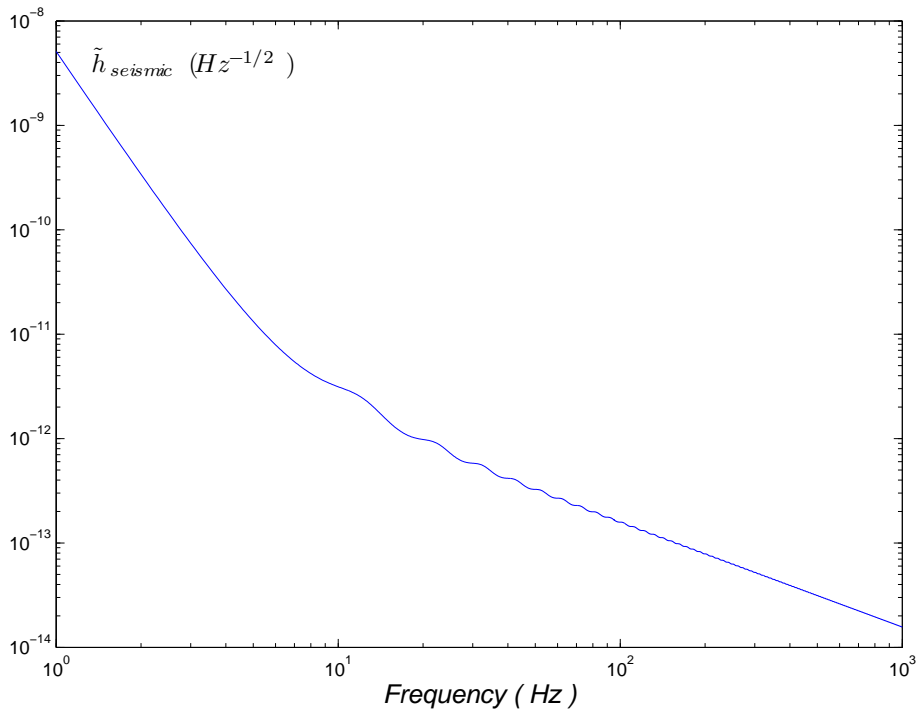}
\caption{Amplitude spectral density curve for seismic noise}
\label{fig3}
\end{figure}

Again, using the above scheme parameters and $a=\lambda/2 \sim 400{\rm nm}$, we can draw the amplitude spectral density curve, as shown in FIG.\ref{fig3}. Seismic noise should be reduced to be smaller than shot noise, which means
\begin{eqnarray}
\Delta (\mathbf{r}_{21}-\mathbf{r}_{32})\cdot \mathbf{k}_{g} \leq \frac{1}{\sqrt{N_0}}.
\end{eqnarray}
Compare FIG.\ref{fig3} with FIG.\ref{fig2}, we find that we need an isolation factor of about $10^8$ in the $10-1000$Hz range, which has already been achieved in LIGO \cite{ligo2009}.

\section{Conclusions}
Through this preliminary study, we find that in high frequency regime the phase shift in our proposed scheme is
proportional to the time derivative of the gravitational wave amplitude. This is a good feature of our scheme, which makes it suitable for detecting gravitational waves in the frequency band $10-1000$Hz, in which there exist a lot of interesting astrophysical sources. Considered current technologies and their future improvements, it is possible for our proposed detector to reach a high sensitivity at a much lower cost. For all these good features, our scheme is a good candidate scheme for building future gravitational-wave detectors.

\begin{center}
\large{{\bf Acknowledgements}}
\end{center}

This work was supported by the National Basic Research Program of China under Grant No.2010CB832805, by the National Natural Science Foundation of China under Grant No.10827404, and also by funds from the Chinese Academy of Sciences.

\end{document}